**Evolution of the domain topology in a ferroelectric**

S. C. Chae[1], Y. Horibe[1], D. Y. Jeong[2], N. Lee[1], K. Iida[3], M. Tanimura[3] and S.-W. Cheong[1*]

[1]Rutgers Center for Emergent Materials and Department of Physics and Astronomy, Rutgers, The State University of New Jersey, Piscataway, NJ 08854

[2]Department of Mathematics, Soongsil University, Seoul 156-743, Korea

[3]Research Department, Nissan Arc, Ltd., Yokosuka, Kanagawa 237-0061, Japan

[*]e-mail: sangc@physics.rutgers.edu

**Topological materials, including topological insulators, magnets with Skyrmions and ferroelectrics with topological vortices, have recently attracted phenomenal attention in the materials science community. Complex patterns of ferroelectric domains in hexagonal $REMnO_3$ (RE: rare earths) turn out to be associated with the macroscopic emergence of $Z_2 \times Z_3$ symmetry. The results of our depth profiling of crystals with a self-poling tendency near surfaces reveal that the partial dislocation (i.e., wall-wall) interaction, not the interaction between vortices and antivortices, is primarily responsible for topological condensation through the macroscopic breaking of the $Z_2$-symmetry.**

Symmetries govern nature ubiquitously from the beauty of human faces [1] to the local gauge invariance of quantum field theory [2]. The spontaneous breaking of symmetry by a variable such as temperature gives rise to a phase transition. Dislocations are common topological defects in materials, which occur during symmetry breaking, and often effectively determine important fundamental crystal properties such as hardness and fatigue behavior, grain boundary development, and charge density wave discommensuration [3-5]. The Burgers vector characterizes each dislocation, and dislocation and anti-dislocation refer to two dislocations with oppositely directed Burgers vectors. Dislocations with Burgers vectors that are not translation vectors with integer times of the underlying lattice unit are called partial dislocations. For example, a charge density wave (CDW) discommensuration can be considered as a partial dislocation with a Burgers vector that is a fraction of a unit cell vector and a few of these discommensurations terminate at a full "CDW dislocation", corresponding to a topological defect with a unit-cell Burgers vector [6, 7]. Dislocations can often interact with each other like particles in a dilute gas [8]. The overlap between the strain fields of adjacent dislocations can induce a paired interaction between the dislocations.

Ferroelectric hexagonal-$REMnO_3$ (RE: rare earths) exhibit intriguing topological defects induced through a trimerization-type structural phase transition [9-12]. This structural transition leads to three structural antiphase domains (α, β, γ), each of which can support either of two directions (+,–) of ferroelectric polarization [13, 14]. The six interlocked structural antiphase and ferroelectric domains of $REMnO_3$ meet in a cloverleaf arrangement that cycles through all six domain configurations [15, 16]. Occurring in pairs, the cloverleaves can be viewed as vortices and antivortices with opposite cycles of domain configuration. We have observed two topologically-distinct types of large-scale vortex/antivortex domain patterns; type-I without any preferred polarization direction, and type-II with a preferred polarization direction [17]. However, the physical nature of switching between type-I to type-II patterns has not been understood.

Herein, we report depth profiling of the ferroelectric domain patterns in two hexagonal $ErMnO_3$ crystal and the symmetry change of the patterns with increasing depth. We have prepared one crystal (EMO-A) with upward polarization favored near the top a-b surfaces and the other crystal (EMO-B) with the opposite tendency (see the detailed experimental methods in the Supplementary Information). The evolution of ferroelectric domain configurations along the c axis was investigated by sequential selective chemical etching and taking optical microscope and atomic force microscope (AFM) images of both of the a-b surfaces of EMO-A. Cross-sectional TEM experiments were performed on EMO-B. Note that the evolution of the domain pattern with increasing depth is not due to the increasing degree of chemical etching and the as discussed in the Supplementary Information section 1.

We have found that the ferroelectric domain configurations at both of the original surfaces of the EMO-A sample were type-II, but became type-I in the interior of the crystal as shown in Fig. 1 (and Fig. S2 in the Supplementary Information section 2). Note that the two parallel surfaces of the crystal favor opposite polarization domains as shown schematically in Fig. S2(e) and S2(h). The differential chemical etching between upward and downward polarization domains resulted in etched surfaces containing shapes of mountain ridges and valley floors as shown in Fig. 1(a). We emphasize that both surfaces show the similar structure with narrow mountain ridges and broad valley floors. Figures 1(b) and 1(c) show the optical microscope images of the top surface after chemical etching of ~1.4 and 7 μm, respectively. These images demonstrate that a type-II pattern with narrow downward polarization domains near the top surface evolves into a type-I pattern with increasing depth: the ridges of the mountains in Fig. 1(a) reflect the narrow downward polarization domains near the top surface, and the valley floors in Fig. 1(a) exhibit the upward polarization domains inside of EMO-A (see also the Supplementary Information section 2 and 3). The corresponding schematics of ferroelectric domain configurations and their evolution are displayed in Figs. 1(d)-1(f). (Figures 1(d) and 1(f) are the schematics of Figs. 1(b) and 1(c), respectively. Figure 1(e) is drawn from the mid-height contour plot of Fig. 1(a).) As demonstrated in Fig. 1(d), the ferroelectric domain patterns near the original surfaces are type-II, but the patterns are type-I inside of the crystal as Fig. 1(f) shows.

Graph theory is useful to understand the seemingly-irregular patterns of ferroelectric domains in hexagonal REMnO$_3$ [17]. For example, figure 1(f) can be considered as a 6-valent graph where six domain walls always merge at each vortex or antivortex core, and each domain is surrounded by an even number of vortices and antivortices. Each domain can be called as an even-gon graphically since it is surrounded by an even number of vertices (vortices and antivortices). This type-I pattern is $Z_2 \times Z_3$ colorable (see the Supplementary Information section 4) in the sense that all domains can be colored with 2 (dark and light) × 3 (red, blue, green) colors in a way that adjacent domains are colored in different colors (proper-colorable), and, for example, a dark red domain is never surrounded by light red domains. These dark and light colors correspond to upward and downward polarizations. On the other hand, figure 1(d) can be considered as a 3-valent graph where all domains with one of dark or light colors are always two-gons. When these two-gons are considered as lines (or edges), then the 6-valent graph with even-gons can be compactified as a "3-valent graph with even-gons", which is 3-proper-colorable. These 3 colors (red, blue and green) correspond to the 3 structural antiphases.

The physical meaning of this $Z_2 \times Z_3$ coloring is that all domains of any ferroelectric domain pattern forming a 6-valent graph with even-gons can be assigned with α+, α−, β+, β−, γ+, and γ− in the way that, for example, an α+ domain is surrounded only by β− and γ− domains. The type-I patterns exhibit $Z_2 \times Z_3$ symmetry in the sense that the topology of the patterns remains intact with respect to the exchange of (+,-) or (α, β, γ) indices, and the symmetry between + and − is broken in the type-II patterns. In other words, the type-II patterns, which can be considered as 3-valent graphs with even-gons after compactification, show only $Z_3$-symmetry with broken $Z_2$-symmetry. All color schemes in the schematics of Figs. 1(d)-1(f) are consistent with the $Z_2 \times Z_3$ coloring. Note that this symmetry approach for domain patterns, regardless of relevant order parameters and microscopic Hamiltonian, reveals the macroscopic topological configuration of the interlocked domains with structural antiphase and ferroelectric polarization.

Interesting systematics emerge when the $Z_2 \times Z_3$ colors in the schematics of Figs. 1(d) and 1(f) are compared. First, the switching from Fig. 1(f) to Fig. 1(d) through Fig. 1(e) can be considered as a topological condensation through the breaking of the $Z_2$-symmetry in the sense that all dark downward polarization domains become two-gons, with each two-gon connecting one vortex and one antivortex. Then, one can consider the opposite process as topological evaporation through the restoration of $Z_2$-symmetry. (See the Supplementary Information section 2 for topological anti-condensation and anti-evaporation.) We note that during topological (anti-)condensation and (anti-)evaporation, most of the cores of vortices and antivortices are hardly influenced since their locations are nearly fixed. Nevertheless, we have observed the appearance of vortex-antivortex pairs with the low generation rate of less than one pair per $4.2 \times 10^{-4}$ μm$^{-2}$, which are discussed in the Supplementary Information section 5.

Investigation by high-resolution TEM confirms that the structural antiphase relationship across one narrow domain is consistent with the $Z_2 \times Z_3$ coloring; i.e., one structural antiphase domain is surrounded by domains with two other structural antiphases. Figure 2(a) shows the optical microscope image of a type-II vortex-antivortex domain on the surface of the etched EMO-B sample. The dark line in the cross-sectional dark-field TEM image of the purple-line region in Fig. 2(a), shown in the inset of Fig. 2(a), corresponds to the narrow upward polarization domain. The dark contrast in the dark-field TEM image originates from breaking of a Friedel's pair in ferroelectrics, and thus confirms that the narrow domain has a ferroelectric polarization opposite to those of its two neighboring domains. Figure 2(b) displays a high-resolution TEM image and the inverse-fast-Fourier-transform (IFFT) image of neighboring domains. Domain boundaries are shown with hatched yellow lines. The broadening of the boundaries may result from the tilting of the boundaries along the depth direction. The solid sinusoidal curves in Fig. 2(c) are intensity scans of the blue and red rectangular areas in the IFFT image of Fig. 2(b), and also extrapolated from the solid curves to check the phase shift between structural antiphases. It should be noted that the modulation in the IFFT image of β+ could be an artifact due to the interference between the phase components in α− and γ− domains. The periodic sinusoidal curves reflect the superlattice modulations due to the Er distortions along the c axis and the tilting of $MnO_5$ hexahedra in $ErMnO_3$. The presence of the phase shift between the two curves demonstrates that two − ferroelectric domains neighboring the narrow + ferroelectric domain have different structural antiphases. This observation, combined with the fact that structural antiphase domain walls are mutually interlocked with ferroelectric domains, does confirm that all neighboring three domains do have different structural antiphases [13].

The topological evolution of a domain pattern with $Z_2$-symmtry breaking with little change of the overall vortex-core structure is primarily associated with the local interaction between the partial dislocations (i.e., the structural antiphase/ferroelectric domain walls), but not with the interaction between vortices and antivortiecs. As shown in Figs. 3(a) and 3(b), the [α−, β+] structural antiphase wall can be considered as a partial dislocation with the Burgers vector of (+, 2π/3), where 2π/3 denotes the phase shift between two structural antiphase domains and + represents the change in polarization direction from − to +. In the same manner, the [β+, γ−], [γ−, β+], and [β+, α−] walls can be considered as partial dislocations with the Burgers vectors of (−, 2π/3), (+,−2π/3), and (−,−2π/3), respectively. If the [α−, β+] and [β+, γ−] walls with the same sign of the Burgers vectors merge, then the resultant wall would be coupled with the Burgers vector of (0, 4π/3), associated with a structural antiphase shift of 4π/3 without changing polarization direction. However, we did not observe any structural antiphase shift without changing polarization direction in our TEM results. This experimental finding, combined with the presence of $Z_2 \times Z_3$ coloring indicates that any wall with the Burgers vector of (0, 4π/3) does not exist. On the other hand, when the [α−, β+] and [β+, α−] walls with Burgers vectors with the opposite sign merge, the Burgers vector becomes zero, i.e., the resultant wall disappears. These results are consistent with the general behavior that two dislocations (or anti-dislocations) with the same Burgers vector tend to be repulsive to each other, whereas a pair of a dislocation and an anti-dislocation with the opposite Burgers vectors can exhibit an attractive interaction [18].

The repulsive interaction between partial dislocations appears responsible for the distribution of wall angles at vortex cores, where partial dislocations of [α−, β+], [β+, γ−], [γ−, α+], [α+, β−], [β−, γ+], and [γ+,α−] meet. As shown in Fig. 3(c), the median of the wall angle distribution of a type-I pattern is slightly below 60° degree. In contrast, the "random" distribution of six different angles should be monotonic as shown in the Supplementary Information section 6. Near vortex cores where partial dislocation walls are proximate to each other, the repulsive interaction between adjacent partial dislocations results in the depression of low angle density with a median value close to 60° in the distribution of six different angles. Note that the average value of the wall angle in our case as well as the random case is ~60°, as expected.

The interactions among partial dislocations and anti-dislocations play an essential role in the processes of topological (anti-)condensation and (anti-)evaporation. In general, the planar structure of partial (anti-)dislocations is associated with a 1/r-type distance dependence of mutual interaction [8, 18]. For example, the partial dislocation pair of [α−, β+] and [β+, γ−] walls with the same Burgers vector can have 1/r-type repulsive interaction. In the presence of electric fields (or effective electric fields in the case of self-poling), favoring − domains, the pair can be stabilized at a short distance where the total potential is minimal. This stabilization appears to be responsible for the presence of narrow domains in the type-II patterns. On the other hand, the partial dislocation-antidislocation pair of [α−, β+] and [β+, α−] walls with the opposite Burgers vectors can have attractive interaction, and be eventually mutually annihilated. Note that a recent electronic structure calculation reported a negligible interaction between domain walls [19]. Further investigation of the interaction is highly needed, considering that self-poling results in the finite width of narrow domains in the type-II domain patterns, which is a hallmark for the existence of a short-range repulsive interaction between partial dislocation pairs, as further discussed below.

This local interaction between partial dislocations governs the macroscopic behaviors of topological $Z_2 \times Z_3$ symmetry and $Z_2$-symmetry breaking. As shown in Fig. 4, the height profile of the AFM image of the middle region of the white-dashed-line rectangle in Fig. 1(c) demonstrates this annihilation process through topological evaporation. First, the narrow two-gon domains in Fig. 1(b) are due to the stabilization of repulsive partial (anti-)dislocation pairs. The topological evaporation process can be visualized from the evolution of red dashed lines from Figs. 4(b) to 4(f), which plot the equal height contour lines of the AFM image. Recall that narrow domains in EMO-A for Fig. 1, unlike EMO-B, are associated with − polarization. Basically, through topological evaporation, α− domains enlarge from narrow two-gons, and the partial dislocation-antidislocation pair of [α−, β+] and [β+, α−] walls can be eventually annihilated, so the β+ domain disappears and the α− domain becomes significantly extended. A similar annihilation process occurs for the partial dislocation-antidislocation pair of [α−, γ+] and [γ+, α−] walls in Fig. 4. In the processes of topological condensation and evaporation shown in Fig. 4, the creation and annihilation of partial dislocation pairs, probably associated with their mutual interaction, are responsible locally for the overall topology change between the type-I and type-II domain patterns.

In summary, we found that the $Z_2 \times Z_3$ symmetry emerges in the seemingly irregular ferroelectric domain patterns of $ErMnO_3$. Poling or self-poling processes induce topological transitions of ferroelectric domains through topological condensation and evaporation. These transitions are associated with the breaking and restoring of the $Z_2$ part of the $Z_2 \times Z_3$ symmetry. The creation and annihilation of pairs of partial dislocations and anti-dislocations with opposite Burgers vectors, i.e., the ferroelectric domain walls interlocked with structural antiphase, and the short-range repulsive interaction between dislocation (or anti-dislocation) pairs are locally responsible for the topological transitions.

**Acknowledgements**

We thank P. Leath for critical reading of the manuscript. This work was supported by National Science Foundation DMR-1104484.

**Figure legends**

FIG. 1 (color online) Depth profiles using sequential chemical etching and $Z_2 \times Z_3$ coloring. (a) Three- dimensional atomic force microscope (AFM) image of the top (001) surface of the EMO-A sample after 7 µm chemical etching. (b) and (c) Optical microscope images of the top (001) surface of EMO-A after 1.4 and 7 µm chemical etching, respectively. Dashed rectangles in Figs. 1(b) and 1(c) correspond to the AFM scanned region of Fig. 1(a). (d) and (f) Schematics of the white-dashed-line rectangle region in Figs.1(b) and 1(c) with $Z_2 \times Z_3$ coloring, respectively. (e) Schematic of an intermediate domain pattern between Figs. 1(d) and 1(f). The depth was estimated from the mid height contour plot of Fig. 1(a).

FIG. 2 (color online) (a) Planar optical microscope image of the type-II pattern of EMO-B after chemical etching. The inset shows a cross-sectional TEM image of the line (purple) region on $\beta+$ domain. (b) High-resolution TEM image of the rectangle (orange) region in the inset of Fig. (a). (c) The structural antiphase shift between $\alpha-$ and $\gamma-$ phases. Upper (red) and lower (blue) sinusoidal waves indicate the superlattice modulations in the $\alpha-$ (red) and $\gamma-$ (blue) phases in Fig. 2(b), respectively.

FIG. 3 (color online) (a) and (b) The local lattice distortions near the $\alpha-/\beta+/\gamma-$ and $\alpha-/\beta+/\alpha-$ domain boundaries in hexagonal $REMnO_3$, respectively. The circles (yellow) with dots (●) or crosses (×) represent the Y ions, where the dots or crosses indicates the upward- or downward- displacements, respectively. The circles at the corners of thick (green) triangles represent the Mn and O ions. The top (light blue) and bottom (dark blue) circles are apical and bottom oxygens of MnO$_5$ hexahedra, respectively. The middle (brown) circles are the Mn ions. The small solid triangles (▲) in the Mn and O ions indicate the directions of ionic displacements. The triangles with thick bars (green) correspond to the Mn trimers. (c) The experimental distribution of the relative angle between adjacent partial dislocations near vortex cores. The red dashed line is drawn as a guide for eyes. The inset shows an AFM image with the definition of the angle (θ) between two adjacent domain boundaries. The dashed black line indicates the average value of 60°, and the arrow indicates the median (~55°) of the angle distribution.

FIG. 4 (color online) Topological evaporation from type-II to type-I. (a) AFM image of the middle region of the white-dashed-line rectangle in Fig. 1(c). (b)–(f) Schematics of evolution from type-II to type-I patterns with attractive interaction and eventual annihilation of partial dislocation-antidislocation pairs ([α−, β+]-[β+, α−] walls and [α−, γ+]-[γ+, α−] walls).

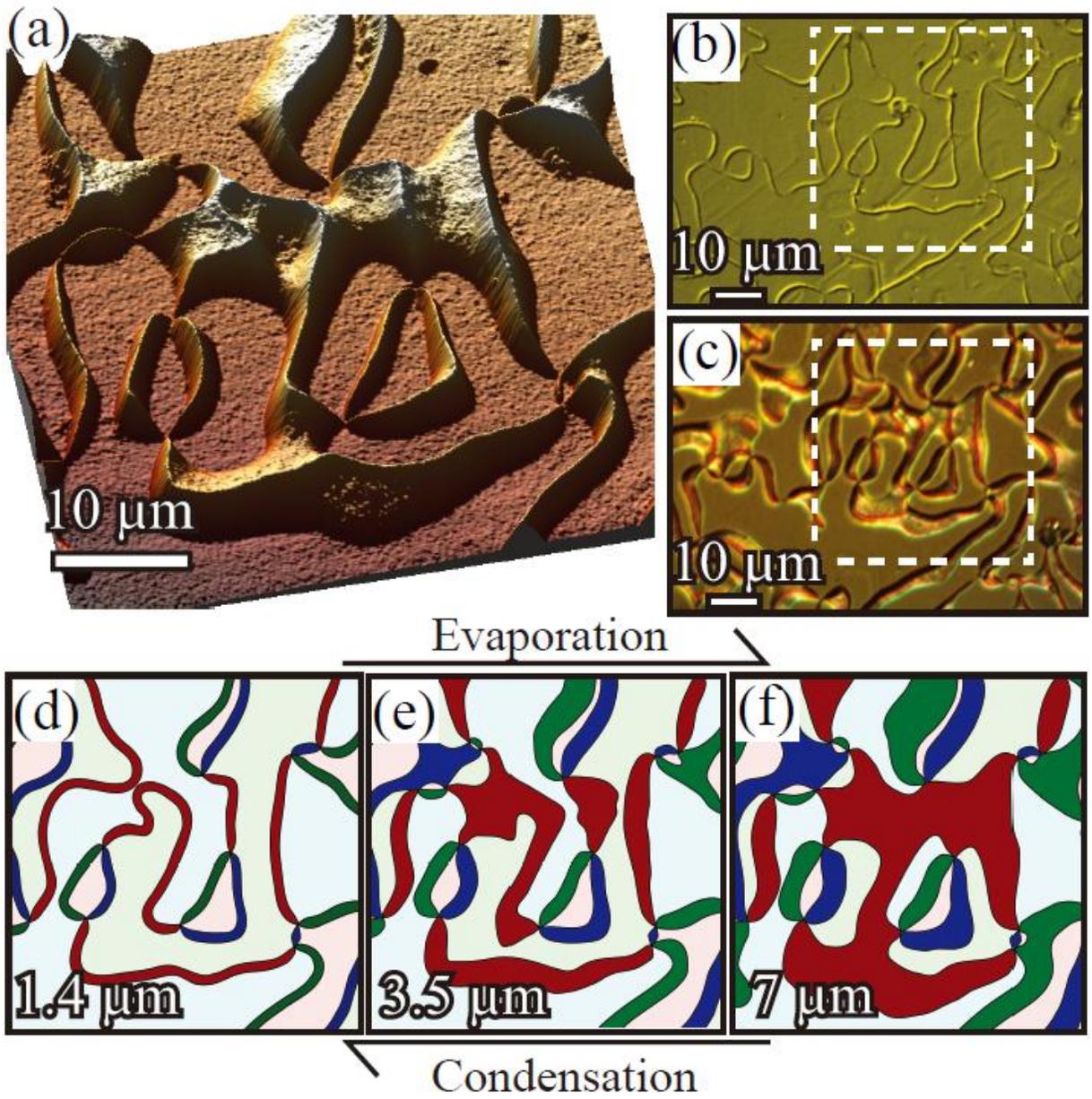

(a)

(b) 10 μm

(c) 10 μm

Evaporation

(d) 1.4 μm

(e) 3.5 μm

(f) 7 μm

Condensation

Figure 1

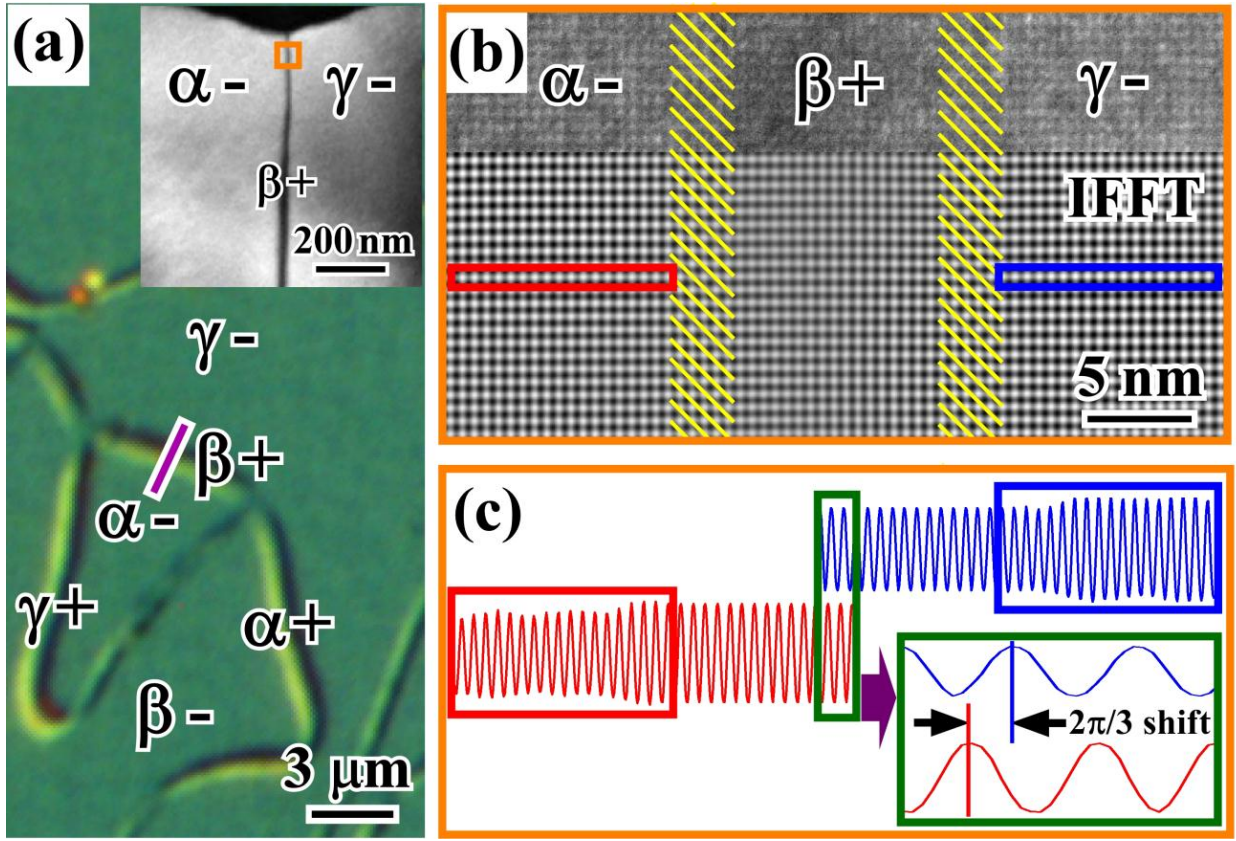

Figure 2

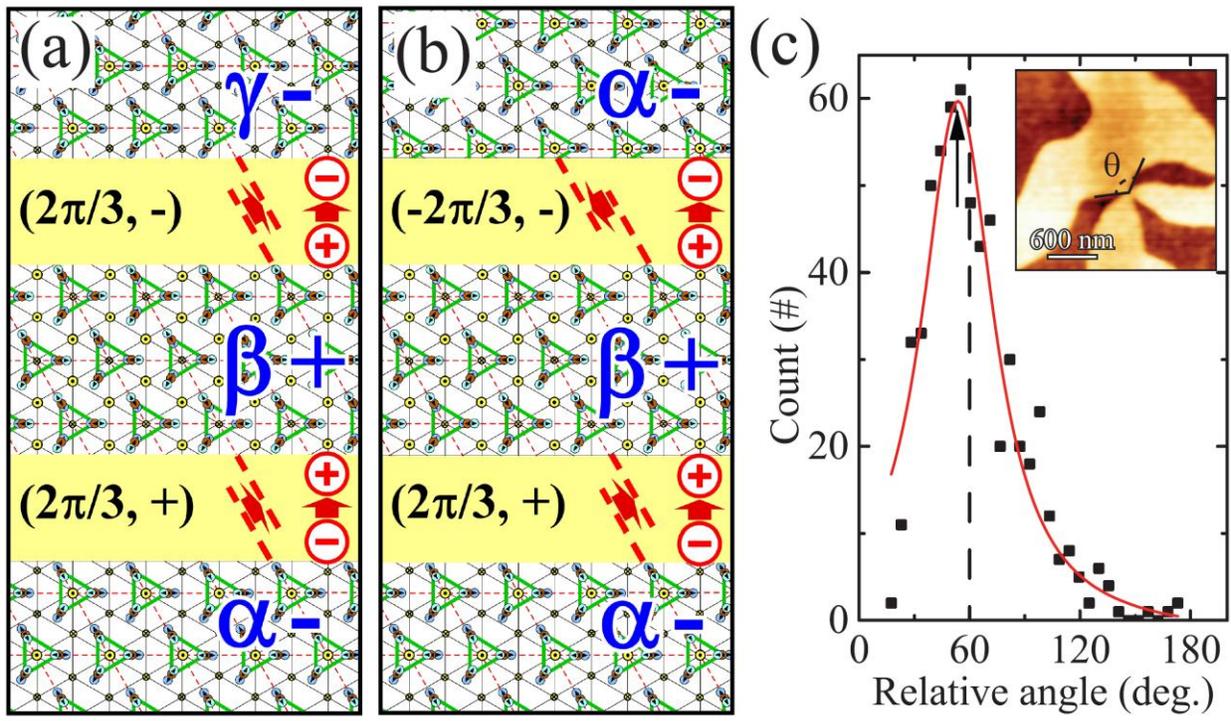

Figure 3

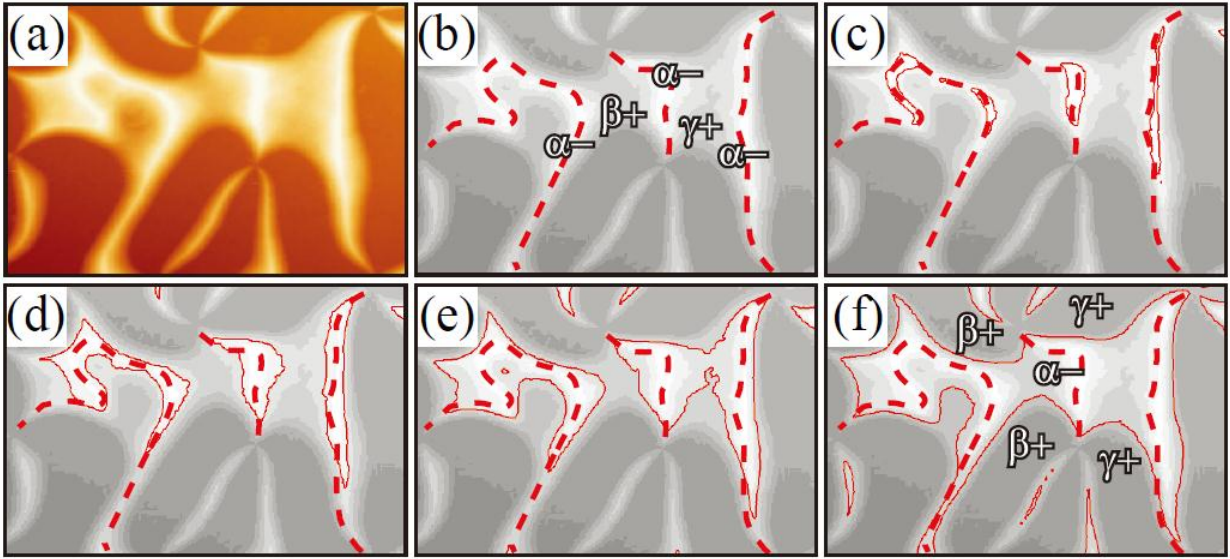

Figure 4

**Evolution of the domain topology in a ferroelectric**

S. C. Chae, Y. Horibe, D. Y. Jeong, N. Lee, K. Iida, M. Tanimura and S.-W. Cheong

Supplementary Information

**Experimental Methods**

ErMnO$_3$ and YMnO$_3$ single crystals were grown using a Bi$_2$O$_3$-based flux method. Two plate-like ErMnO$_3$ single crystals (EMO-A and EMO-B; ~1×0.5×0.02 mm$^3$, thin along the c axis) were cooled slowly from 1200 $^o$C to induce a vortex-antivortex pattern, and then annealed at 850 $^o$C in various gas environments in order to induce self-poling effects [1]. EMO-A shows wide upward polarization domains and narrow downward polarization domains near the top a-b surfaces, but EMO-B exhibits the opposite tendency. In order to observe the ferroelectric domain configurations, plate-like ErMnO$_3$ crystals were etched chemically in phosphoric acid at 130 $^o$C. Note that phosphoric acid selectively etches the surface of upward polarization domains, and EMO-B-type crystals are more common, so the EMO-A sample was specifically prepared to perform depth profiling by acid etching. The atomic force microscopy (AFM) experiments were performed using a Nanoscope IIIA (Veeco). All optical and AFM images were taken at room temperature. The specimens for cross-sectional transmission-electron-microscopy (TEM) experiments were prepared with a focused-ion beam (FIB) technique. Our TEM studies were carried out using JEOL-2010F and Hitachi H-9000 operated at 200 kV and 300 kV, respectively. The crystal orientation and reflection spots were indexed using the hexagonal notation with the space group of P6$_3$cm. For the electric poling experiment, Ag electrodes were attached to an YMnO$_3$ crystal. After electric poling at low temperature with 77K, the Ag electrodes were removed mechanically.

## Section 1. Domain pattern change from type-I to type-II by external electric poling

In order to check the external electric poling on the domain pattern change, we applied the external electric field which is larger than coercive field of YMnO$_3$ crystal. For the electric poling experiment, Ag electrodes were attached to an YMnO$_3$ crystal. After electric poling at low temperature with 77K, the Ag electrodes were removed mechanically. Figure S1 shows the atomic force microscope images of negatively poled surface, i.e., negative bias was applied to this surface. The atomic force microscope images were observed after chemical etching. The electric poling in poled region leaves preferred wide upward domain pattern and the narrow downward polarization, whereas the unpoled region shows clear the type-I pattern. It was reported that the upward polarization domain etched faster than the downward polarization domain [2].

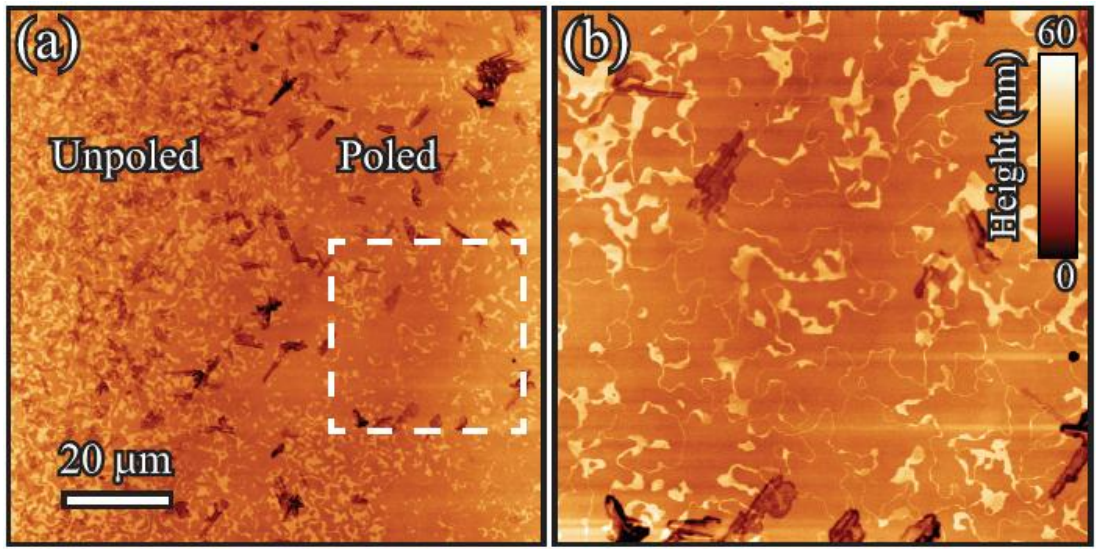

FIG. S1 The atomic force microscope scan images of a (001) YMnO$_3$ surface, a part of which was poled by applying electric fields. (a) AFM scan image near the boundary region of an Ag electrode that was used to apply electric fields. The Ag electrode was removed mechanically after electric poling. (b) Enlarged view of the white dashed rectangle region of Fig. S1(a). The type-II vortex pattern with topological condensation is evident in the poled region, compared with a type-I pattern in the unpoled region.

## Section 2. Topological (anti-)condensation and (anti-)evaporation through the breaking and restoring of $Z_2$-symmetry, respectively.

The ferroelectric domain configurations at both of the top and bottom surfaces of EMO-A are revealed using differential chemical etching between upward and downward polarization domains. Figures S2(a) and S2(b) are identical with Fig. 1(b) and 1(c). Figures S2(c) and S2(d) show the optical microscope images of the bottom surface after 30- and 150-minutes chemical etching, respectively. The corresponding schematics of ferroelectric domains are displayed in Figs. S2(e)-S2(h).

As chemical etching progresses, the type-II patterns at both of the surfaces turn into type-I patterns without any preferred polarization direction. These domain pattern changes can be understood in terms of breaking and restoring of the $Z_2$ part of the $Z_2 \times Z_3$ symmetry in type-I patterns. As discussed in the main text, for one fixed reference frame along the positive [001] direction, the top surface shows the process from Fig. S2(a) to Fig. S2(b) and vice versa through topological evaporation and condensation, respectively. The topology of the dark downward-polarization domains changes through these processes. The bottom surface shows the domain pattern change from Fig. S2(d) to Fig. S2(c) and vice versa, and these domain pattern changes at the bottom surface are analogue with topological condensation and evaporation explained above. However, these changes are associated with the topology change of the "bright" upward polarization domains. So, these processes at the bottom surface can be considered as topological anti-condensation and anti-evaporation.

We emphasize the enormous power of graph theory to analyze and understand the complex domain patterns that we have observed. (The detailed graph-theoretical proof of $Z_2 \times Z_3$ symmetry and coloring is discussed in the section S4.) The identification of the topological condensation process accompanying the $Z_2$-symmetry breaking is possible only through graph theoretical analysis. The entire concept of topological condensation and evaporation is not possible without graph theoretical consideration of the large-scale patterns. In addition, our system is the first case where graph theory is applied to understand the non-trivial connectivity of large-scale complex domain patterns in functional solid-state materials. Furthermore, the fact that graph theoretical coloring works for an entire complex domain pattern unveils the highly-nontrivial aspect that each domain wall in the pattern is an interlocked boundary with a structural antiphase domain wall and a ferroelectric domain wall, and there exist no non-interlocked or disordered walls in the pattern with numerous walls. Finally, we note that utilizing the $Z_2 \times Z_3$ coloring, we can readily identify the polarization directions and trimerization phases of all of millions of domains if we know the exact nature of only "two" neighboring domains

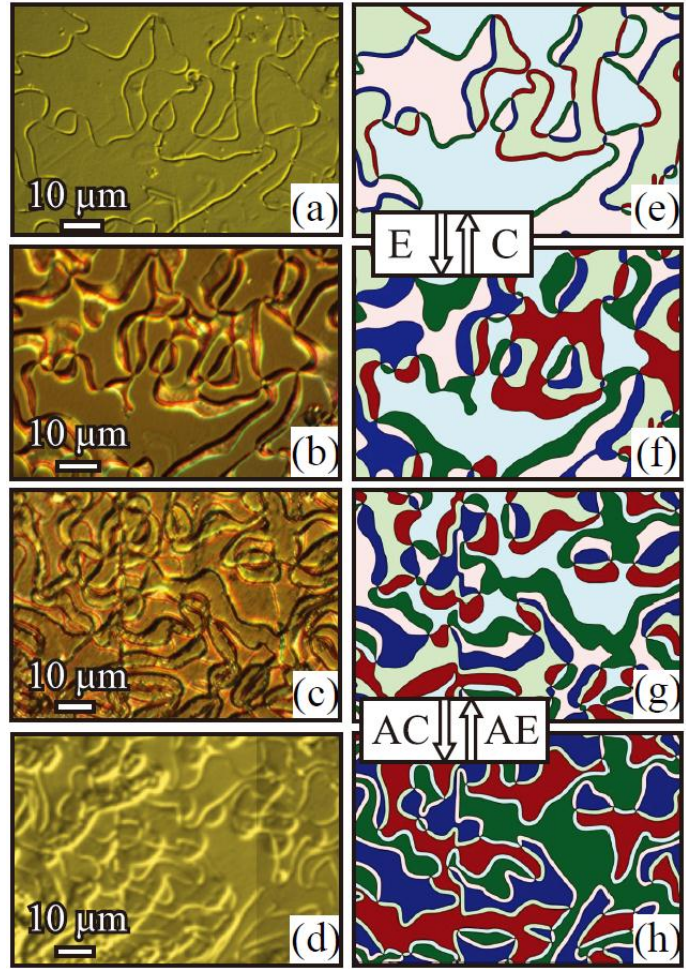

FIG. S2 Depth profiles using sequential chemical etching and $Z_2 \times Z_3$ coloring. (a) and (b) Optical microscope image of the top (001) surface of EMO-A after 30 and 150 minutes chemical etching, respectively. (c) and (d) Optical microscope images of the bottom ($00\bar{1}$) surface of EMO-A after 150 and 30 minutes chemical etching, respectively. (e)-(h) Schematics of Figs. S2(a)-2(d) with $Z_2 \times Z_3$ coloring, respectively. The (001) and ($00\bar{1}$) surfaces show the type-II patterns favoring upward polarization and downward polarization with respect to the [001] axis, respectively. However, the heavily etched (001) and ($00\bar{1}$) surfaces, i.e., inside regions of the crystal exhibit the type-I pattern without any preferred polarization direction. Thus, $Z_2 \times Z_3$ symmetry exists inside regions, but the $Z_2$ part of $Z_2 \times Z_3$ symmetry is broken at surfaces. The breaking and restoration of $Z_2$-symmetry are associated with the (anti-)condensation and (anti-)evaporation as depicted with arrows with (A)C and(A)E in Figs. S2(e)-S2(h), respectively.

**Section 3. Sequential optical microscope images against chemical etching**

The evolution of a vortex domain pattern was monitored with increasing chemical etching time. As shown in Fig. S3, the overall domain pattern change was occurred within 30 minutes etching; the domain pattern transition from type II to type I is observed from Fig. S3(b) to Fig. S3(c). The optical microscope image, Fig. S3(f), after additional chemical etching with 1hour doesn't show much change with respect to Fig. S3(c). This sudden change is consistent with the phase transition nature of topological condensation and evaporation, and is not consistent with an effect induced continually with increasing etching.

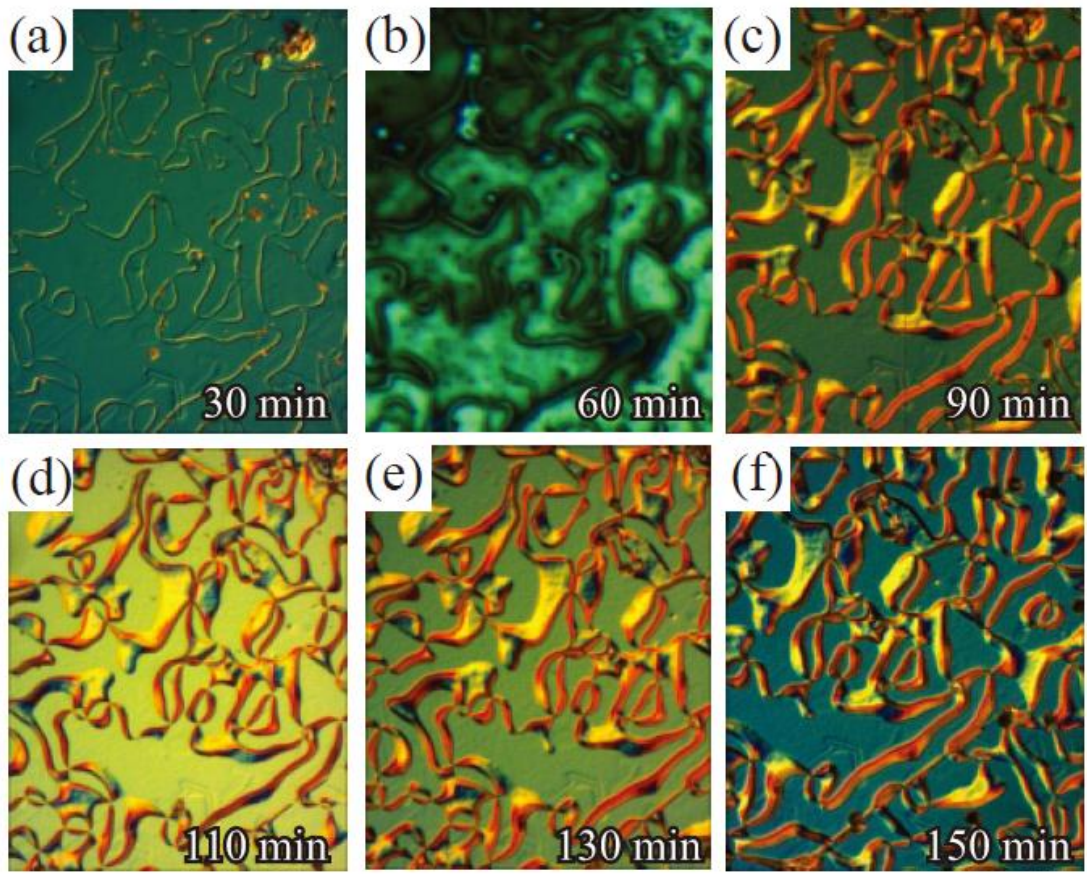

FIG. S3 Evolution of a vortex domain pattern with increasing chemical etching time. (a)-(f) Optical microscope images after etching for 30, 60, 90, 110, 130 and 150 minutes, respectively. The sudden change from type-I to type-II patterns at 60-90 minutes is consistent with the phase transition nature of topological condensation and evaporation.

**Section 4. Mathematical proof of $Z_2 \times Z_3$ coloring for 6-valent graphs with even-gons.**

We have proven mathematically that any 6-valent graph with even-gons is $Z_2 \times Z_3$-colorable All faces of any 6-valent graph with even-gons are 2-proper-colorable. After this 2-proper-coloring, faces with each color can be further colored with 3 colors in the way that each face is never surrounded by any face with the same "second" color. This process turns out to be unique.

## 1. Preamble

We say $G$ is a graph if it consists of a nonempty set $V(G)$ of vertices together with a set $E(G)$ of edges which joins two vertices. If there is no confusion in context, we use $V$ and $E$ instead of $V(G)$ and $E(G)$, respectively. The number of edges incident to a vertex $u$ is called the degree of the vertex $u$ and is denoted by $d(u)$. A graph is called $r$-valent if $d(u)=r$ for all $u \in$ V. A graph is called planar if all the edges can be redrawn on the plane without any intersection except for vertices. A region surrounded by the edges in a planar graph is called a face, and the face which is surrounded by $k$ edges is called a $k$-gon. We usually call two-gon a digon. Two vertices are said to be adjacent if they joined by an edge. Two edges (faces, respectively) are adjacent if they share a common vertex (common edge, respectively).

## 2. Coloring graphs

Let $G$ be a planar graph. We can color vertices, edges, and faces and call them a vertex coloring, an edge coloring, and a face coloring, respectively. A vertex (edge, face, respectively) coloring is called proper if adjacent vertices (edges, faces, respectively) are in different colors. A graph is called $k$-vertex (edge, face, respectively) colorable if $k$ is the smallest positive integer which yields a proper vertex (edge, face, respectively) coloring.

We have several known results about the coloring:

**Theorem 1** Every planar graph whose faces are all even gons is 2-vertex colorable.

**Theorem 2** A graph is bipartite if and only if it is 2-vertex colorable.

## 3. Main results

In this section, we want to show that every 6-valent planar graph whose faces are all even-gons has a so-called $Z_2 \times Z_3$ coloring. Let us first define $Z_2 \times Z_3$ coloring. Suppose $G$ is a 6-valent planar graph whose faces are all even-gons(including digons). From Theorem 1 and 2 above, we know that $G$ is bipartite and 2-vertex colorable. Also, it is 2-face colorable because the dual of $G$ is also a planar graph whose faces are all 6-gons. Hence we can color the faces of $G$ properly with orange and blue. However, our interest is to color the faces of $G$ in a special way as follows:

1. Every face is properly colored in 6 different colors: $B_1$, $B_2$, $B_3$, $W_1$, $W_2$, and $W_3$.

2. Every face colored by $B_i$, $i$=1, 2, and 3 is surrounded by the faces of alternating colors $W_j$ and $W_k$, ($j \neq i$ and $k \neq i$).

3. Every face colored by $W_i$, $i$=1, 2, and 3 is surrounded by faces of alternating colors $B_j$ and $B_k$, ($j \neq i$ and $k \neq i$).

We call such type of coloring described above as **$Z_2 \times Z_3$ face coloring**.

The following is the main theorem.

**Theorem 3** Let $G$ be a 6-valent planar graph whose faces are all even-gons including digons (two-gons). Then $G$ has a $Z_2 \times Z_3$ face coloring.

To prove this theorem, we state and prove some lemmas.

Let $G$ be a 6-valent planar graph whose face are all even-gons. Let $F_0$ be a face surrounded by $2m$ edges $e_1$, $e_2$, . . ., $e_{2m}$ for some integer $m$ ($m$>1) where each $e_i$ joins two vertices $v_i$ and $v_{i+1}$ for all $i = 1, 2, . . ., 2m - 1$, and the edge $e_{2m}$ joins vertices $v_{2m}$ and $v_1$. Let $Fi$, $i = 1, 2, . . ., 2m$, be the face which shares the edge $e_i$ with $F_0$ in common (see Fig. S4(a)).

Now, we want to introduce an operation $\alpha$ on the face $F_0$ of $G$ as follows:

Choose an edge and choose another edge which is the second position from the chosen edge. For example, if we choose $e_1$, we may choose either $e_3$ or $e_{2m-1}$. Suppose that we choose $e_3$. Then delete both edges and draw new edge $e'_2$ and e'$_1$, which connect vertices $v_2$, $v_3$ and $v_1$, $v_4$, respectively.  (see Fig. S4(b)).

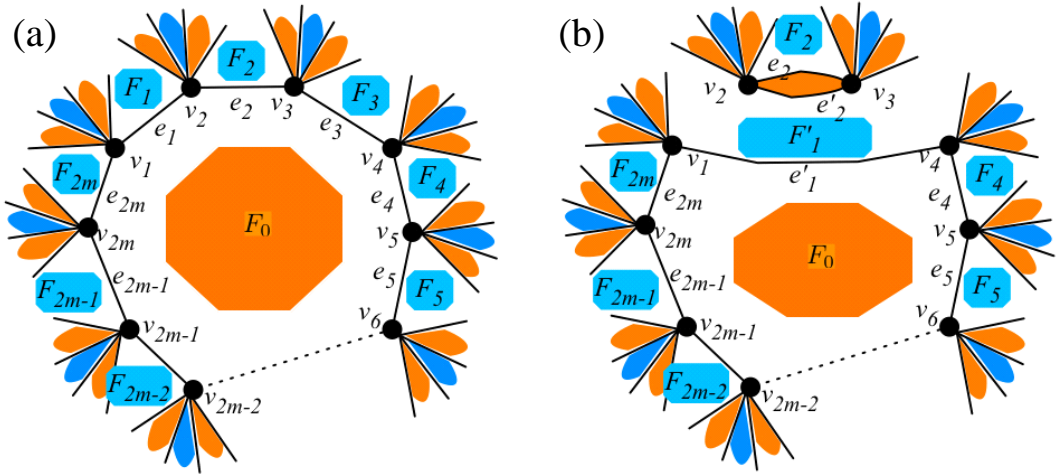

FIG. S4 Schematic diagram of the operation $\alpha$.

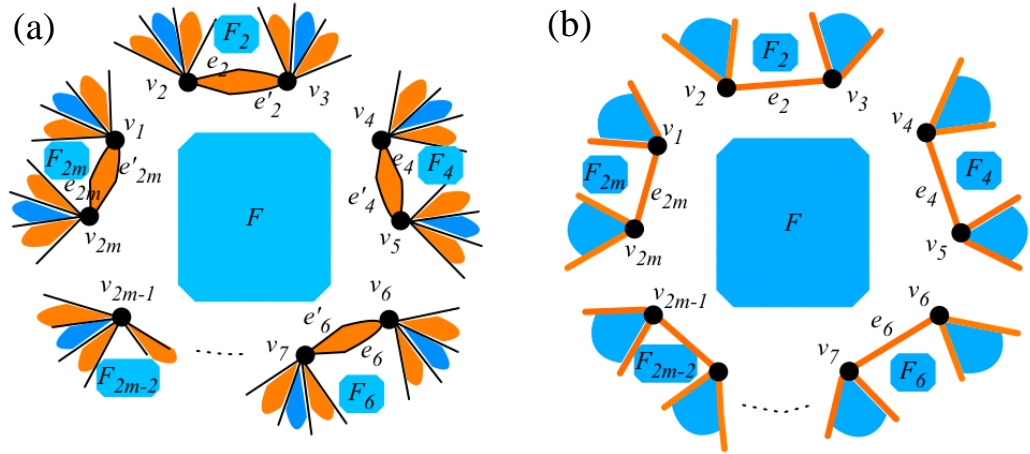

FIG. S5 From 6-valent graph to 3-valent graph.

We get the following results by this operation $\alpha$ on the face $F_0$ of $G$.

i. The $2m$-gon $F_0$ is divided into two faces, one is a $(2m-2)$-gon $F'_0$ and a digon.

ii. The two faces $F_1$ and $F_3$ become one face $F'_1$. Moreover, if $F_1$ is an $s$-gon and $F_3$ is a $t$-gon, then $F'_1$ becomes an $(s+t)$-gon.

Let $G_1$ be the graph obtained by the operation $\alpha$. Clearly, $G_1$ is again a 6-valent planar graph whose faces are all even-gons.

**Lemma 4** Every 6-valent planar graph $G$ whose faces are all even-gons can be transformed to a 3-valent planar graph $G''$ whose faces are all even-gons.

**Proof** Let $G$ be a 6-valent planar graph whose faces are all even-gons. Then $G$ is 2-face colorable. Suppose we color the faces of $G$ properly with orange and blue. Let us choose an orange colored face $F_0$ which is not a digon, and apply the operation $\alpha$ repeatedly until the face $F_0$ splits into small digons of color orange. (see Fig. S5(a)) The resulting graph is still a 6-valent planar graph whose faces are all even-gons. We repeat this process until all the orange colored faces become digons. Then we remove one edge from each orange colored digon so that all the orange colored digons are vanished. (It is like shrinking all orange colored digon to an edge). We finally have a 3-valent graph $G''$ whose faces are all blue colored even-gons. (see Fig. S5(b)) Note that there are odd number of edges between two faces $F_{2k}$ and $F_{2(k+1)}$ (k=1, 2,…,m-1) along the sides of the face F. ∎

**Lemma 5** Every 3-valent planar graph $G$ whose faces are all even-gons is 3-face colorable.

**Proof** Let $G$ be a 3-valent planar graph whose faces are all even-gons. Suppose we color a face with the color $W_3$, then the surrounding faces can be colored in alternation with the colors $W_1$ and $W_2$ as illustrated in Fig. S6(a). Apply this coloring method to the faces one by one. Then all the faces can be colored properly with 3 different colors $W_i$, $i$=1, 2, and 3 in such a way that every face colored by $W_i$, $i$=1, 2, and 3 is surrounded by the faces of alternating colors $W_j$ and $W_k$ ($j$≠$i$ and $k$≠$i$). ∎

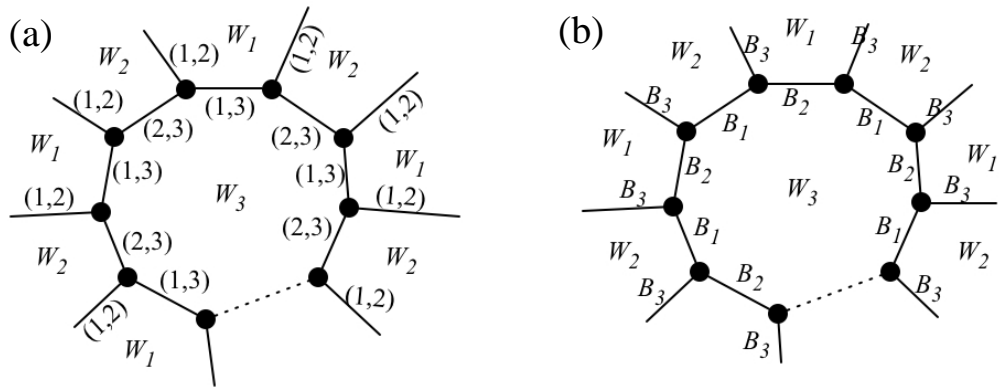

FIG. S6 Schematic diagram of 3-colorings.

**Corollary 6** Every 3-valent planar graph $G$ whose faces are all even-gons is 3-edge colorable.

**Proof** Suppose $G$ is a 3-valent planar graph whose faces are all even-gons. According to Lemma 5, we can color all the faces in 3 different colors $W_i$, $i$=1, 2, and 3 so that every face colored by $W_i$, $i$ = 1, 2, and 3 is surrounded by the faces of alternating colors $W_j$ and $W_k$ ($j{\neq}i$ and $k{\neq}i$). Now, label each edge as $(i, j)$ when it is a common edge of two faces colored by $W_i$ and $W_j$. We see that there are only 3 possibilities; (1, 2), (1, 3), and (2, 3) as shown in Fig. S6(a). For the edge with label $(i, j)$, assign color $B_k$, $k{\neq}i$, $k{\neq}j$, $k$=1, 2, and 3, that is, assign the color $B_1$ ($B_2$, $B_3$, respectively) to the edge whose label is (2, 3) ((1, 3), (1, 2), respectively). (see Fig. S6(b)). Clearly, it is a proper edge coloring. Thus, G is 3-edge colorable. ∎

We are ready to prove the main theorem.

**Theorem 3** Let $G$ be a 6-valent planar graph whose faces are all even-gons including digons (two-gons). Then $G$ has a $Z_2{\times}Z_3$ face coloring.

**Proof** Let $G$ be a 6-valent planar graph whose faces are all even-gons. We can apply the operation $\alpha$ repeatedly to get a 3-valent planar graph $G_1$ whose faces are all even-gons. (Lemma 4). Then the faces of $G_1$ can be colored in 3 different colors $W_i$, $i$=1, 2, and 3 so that each face colored by $W_i$, $i$ = 1, 2, and 3 is surrounded by the faces of alternating colors $W_j$ and $W_k$ ($j{\neq}i$ and $k{\neq}i$) (Lemma 5). Then every edge can be colored with 3 different colors $B_i$, $i$ = 1, 2, and 3 in such a way that every edge of color $B_i$, $i$ = 1, 2, and 3 is between the faces of colors $W_j$ and $W_k$ ($j{\neq}i$ *and $k{\neq}i$*) (Corollary 6). Moreover, each face of color $W_i$, $i$ = 1, 2, and 3 is surrounded by the edges of alternating colors $B_j$ and $B_k$ ($j{\neq}i$ and $k{\neq}i$). Now, we make each edge a digon which inherit the color of the edge. (In other words, inflate each edge to get a digon. (see Fig. S7(a)) And, reverse the repeated operation $\alpha$ to recover the original face. For example, the edges of color $B_2$ become a face of color $B_2$ while the face colored $W_3$ splits into many small faces. Also, the face of color $B_2$ is surrounded by the faces of alternating colors $W_1$ and $W_3$. (see Fig. S7(b)) After reversing the whole application of the operation $\alpha$, we recover the graph G which has the $Z_2{\times}Z_3$ face coloring. ∎

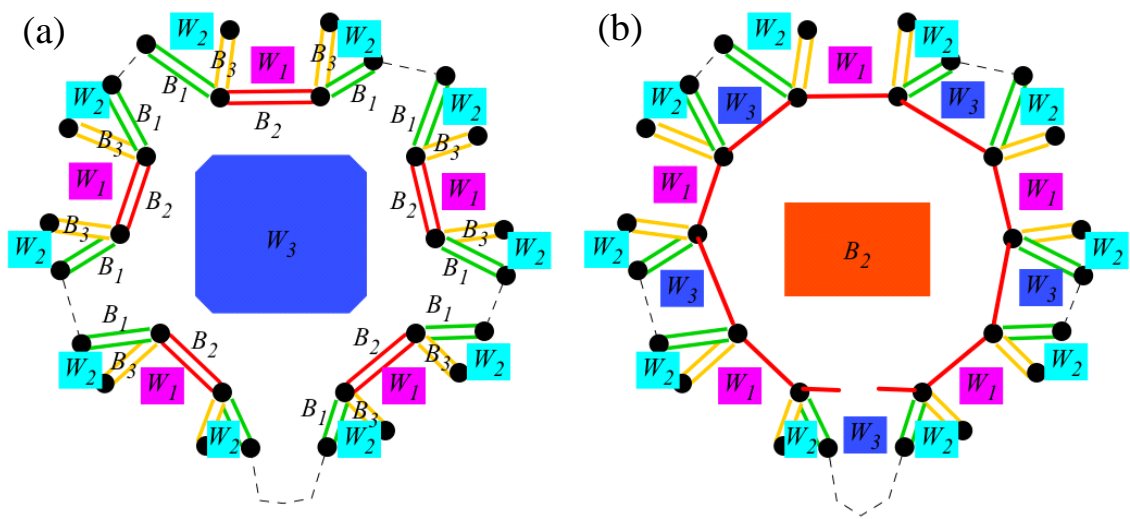

FIG. S7 Schematic diagram of $Z_2 \times Z_3$ face coloring.

**Section 5. The disappearance of vortex-antivortex pairs through topological condensation.**

We have observed the disappearance of vortex-antivortex pairs through topological condensation, as shown in Fig. S8. Figures S8(a) and S8(b), identical with Figs. 1(b) and 1(c), show optical microscope images of the top surface and an inside region of one area, respectively. Figures S8(c) and S8(d) display enlarged views of the rectangle region in Figs. S8(a) and S8(b), respectively. Figures S8(e) and S8(f) show the domain schematics of the area with proper coloring, where two vortex-antivortex pairs disappear through topological condensation. The (dis)appearance of two vortex-antivortex pairs is depicted in Figs. S8(g), S9(a) and S9(b) in a three-dimensional manner. Lowering the height in Figs. S9(a) and S9(b) can be viewed as topological evaporation, and two vortex-antivortex pairs appear during topological evaporation. Topological condensation corresponds to the increase of the height, and each vortex-antivortex pair disappeared during topological condensation. Note that it is not clear if this (dis)appearance of vortex-antivortex pairs is due to self-poling or accidental occurrence through entropical fluctuations.

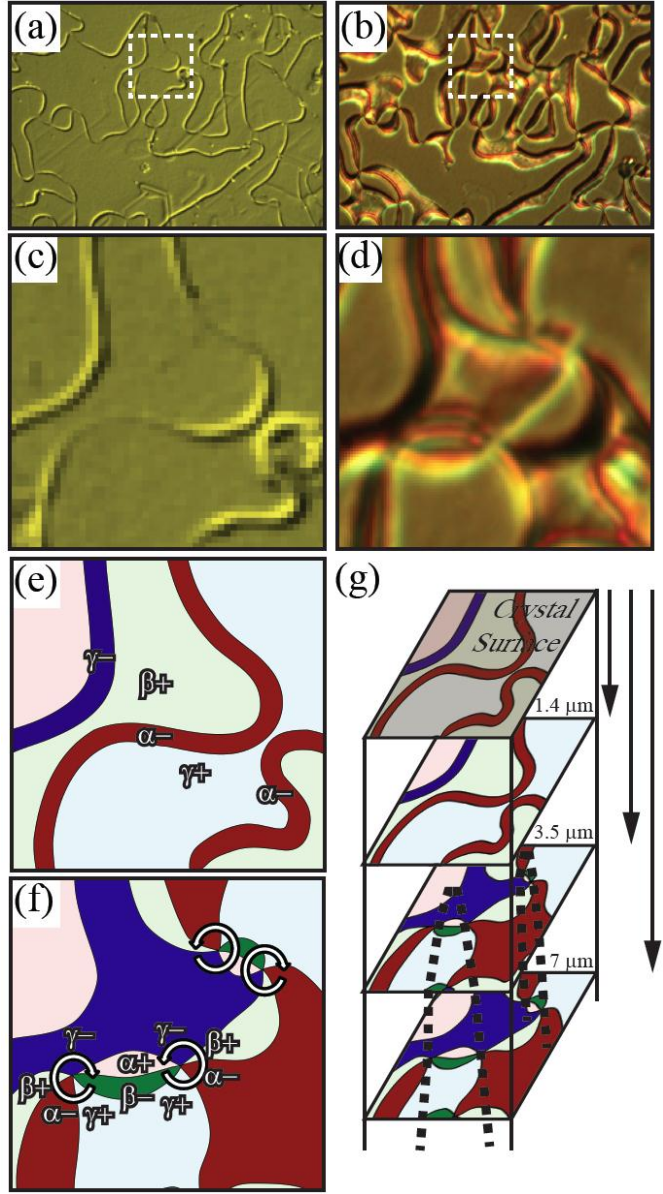

FIG. S8 The appearance of vortex-antivortex pairs during the topological evaporation process. (a) and (b) Optical microscope images identical with Figs. 1(b) and 1(c), respectively. (c) and (d) The enlarged images of the white-dashed-line rectangle region in Figs. S8(a) and S8(b), respectively. (e) and (f) Schematics of Figs. S8(c) and S8(d) with proper coloring, showing the appearance of two vortex-antivortex pairs through topological evaporation. (g) A three-dimensional schematic of the (dis)appearance of vortex-antivortex pairs. The dashed lines indicate vortex cores.

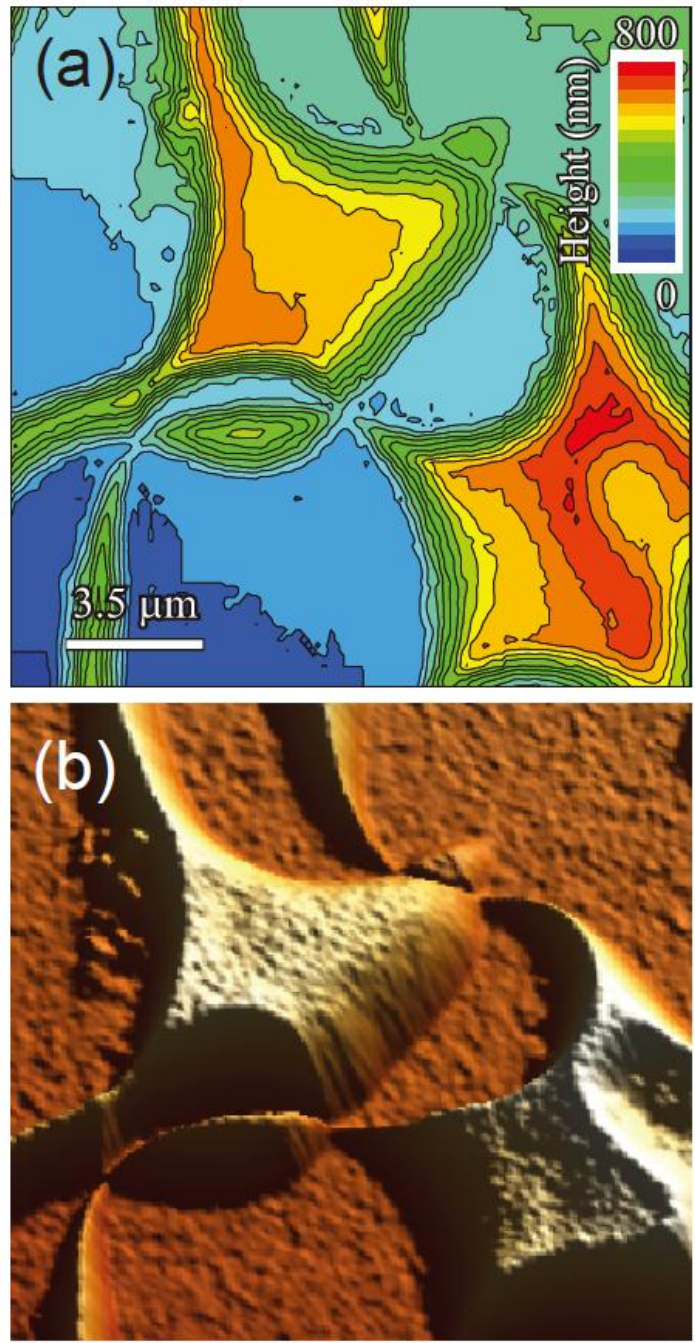

FIG. S9 Three-dimensional view of the appearance process of two vortex-antivortex pairs and topological evaporation. (a) and (b) The equal-height contour plot and three-dimensional view of the atomic force microscope image of the area shown in Fig. S8(d).

**Section 6. The expected angle between partial dislocations near vortex cores without mutual interaction.**

The "random" distribution of the adjacent-two-lines angle of six straight lines hinged at a point is simulated, and the result is monotonic as shown in Fig. S10(b). This result is distinct from the Bell-curve-like distribution of the experimental angle between adjacent partial dislocations near vortex cores displayed in Fig. 3(c). The average angle value for both the simulation with randomness and the experimental distribution is 60°. However, the simulation shows a monotonic decrease of occurrence with increasing angle value, but the low-angle occurrence in the experimental distribution is highly suppressed, and the median for the experimental angles is ~55° as shown in Fig. 3(c). Note that the experimental distribution was estimated from the pattern in Fig. S8(a), obtained using another $ErMnO_3$ crystal that was annealed in air to exhibit type-I patterns on surfaces. These results suggest strongly the existence of the repulsive interaction between partial dislocations.

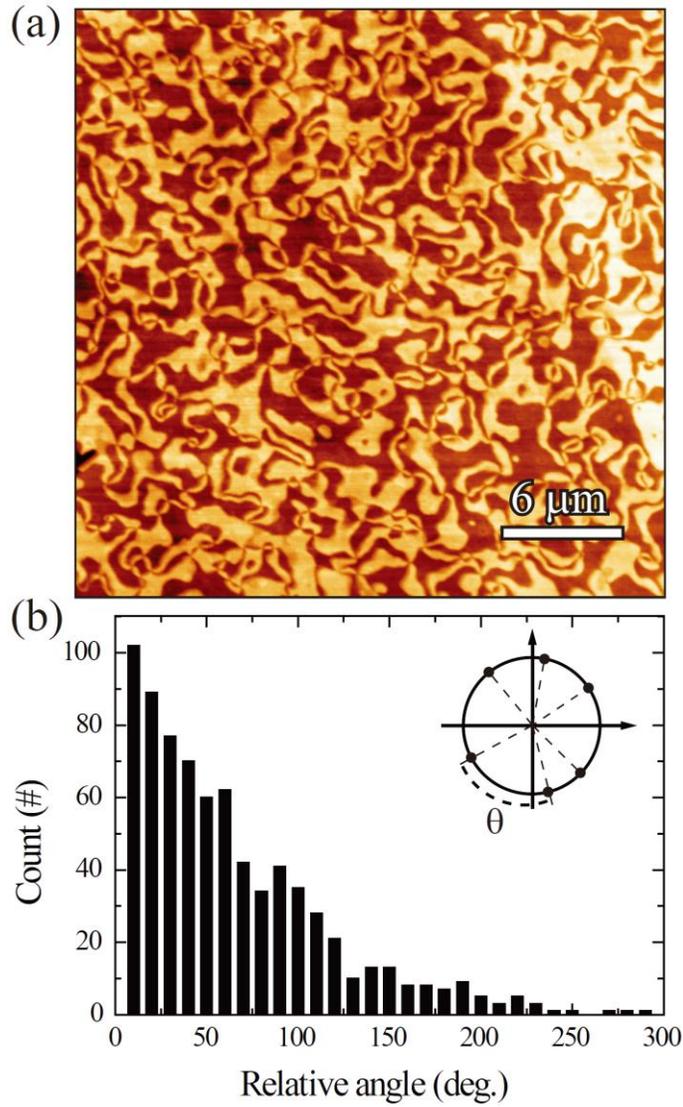

FIG. S10 (a) The atomic force microscope image of the chemically-etched surface of a ErMnO₃ crystal, showing a type-I pattern. This image was used for the analysis of relative angle distribution in Fig. 3(c). (b) The simulated histogram of the angle between adjacent partial dislocations in the random case. The inset shows the schematic of the angle between adjacent partial dislocations for our simulation, and the simulation is performed for the random case, i.e. without any correlation between partial dislocations.